**High-Field, Quasi-Ballistic Transport in Short Carbon Nanotubes**

Ali Javey,[1] Jing Guo,[2] Magnus Paulsson,[2] Qian Wang,[1] David Mann[1], Mark Lundstrom[2] and Hongjie Dai[1]

[1] Department of Chemistry and Laboratory for Advanced Materials, Stanford University, Stanford, CA, 94305

[2] School of Electrical and Computer Engineering, Purdue University, West Lafayette, IN, 47907

Single walled carbon nanotubes with Pd ohmic contacts and lengths ranging from several microns down to 10 nm are investigated by electron transport experiments and theory. The mean free path (mfp) for acoustic phonon scattering is estimated to be $l_{ap}$~300 nm, and that for optical phonon scattering is $l_{op}$~15 nm. Transport through very short (~10 nm) nanotubes is free of significant acoustic and optical phonon scattering and thus ballistic and quasi-ballistic at the low and high bias voltage limits respectively. High currents of up to 70 μA can flow through a short nanotube. Possible mechanisms for the eventual electrical breakdown of short nanotubes at high fields are discussed. The results presented here have important implications to high performance nanotube transistors and interconnects.

Email: hdai@stanford.edu

Recent progress in ohmic contacts to single-walled carbon nanotubes (SWNTs) has facilitated the elucidation of the intrinsic electron transport properties of these novel materials and the advancement of high performance nanotube electronics.[1-6] With high quality as-grown chemical vapor deposition (CVD) materials and ohmic contact strategies, several groups have observed ballistic electron transport in metallic nanotubes,[2,3] and more recently, in semiconducting SWNTs by Javey et al.[5] The hallmark of ballistic transport in SWNTs is the conductance (G) of the samples reaching four quantum units $4e^2/h$ (resistance R ~6.5 kΩ) and the manifestation of phase coherent resonance at low temperatures.[2,3,5,7] However, observations of ballistic transport in SWNTs have been mostly limited at the low bias voltage (< 0.1 V) regime thus far. Yao et al. have shown that in the high bias (> 0.16 V) regime, current saturation at the 25 μA level occurs due to electron backscattering by optical or zone boundary phonon emissions.[1] Similar current saturation has been observed in ohmically contacted semiconducting SWNTs recently.[5] These results hint that for SWNTs longer than the mfp of optical phonon scattering ($l_{op}$ in the range of 10 –100 nm from Ref. 1), high bias electron transport is not truly ballistic.

Here, we investigate ohmically contacted metallic SWNTs with lengths systematically scaled from the micron scale down to 10 nm. Pd metal affords near ohmic contacts to these nanotubes with very high reproducibility,[5,7] a key to this investigation. Theoretical analysis and fitting of experimental data reveal $l_{ap}$ ~300 nm and $l_{op}$ ~ 15 nm. Consequences of these length scales include, first, for metallic SWNTs with length L≪$l_{ap}$, the low bias conductance is essentially ballistic and nearly temperature independent. Second, for nanotubes with L ~ 10 nm, the current carrying capability reaches up to 70 µA, well exceeding the highest (25 µA) reported previously. Short SWNTs are, therefore, close to truly ballistic conductors. Eventual high field breakdowns of these ultrashort nanotubes are also discussed.

Individual SWNT devices were fabricated by patterned CVD growth of SWNTs on $SiO_2$/Si wafers,[8] followed by electron beam lithography (EBL, with a Raith 150 EBL system), metal deposition, and liftoff to form Pd source/drain (S/D) top-contacts. The lengths of SWNTs were defined by the distances between the edges of the Pd electrodes, and varied down to 10 nm (Fig. 1). The diameters of the nanotubes used for this work were all in the range of 1.5 to 2.5 nm. The heavily p-doped Si substrate was used as the back-gate with thermally grown $SiO_2$ as the gate dielectric. All room temperature data presented here were recorded with the samples exposed to air.

Nearly all of our Pd contacted metallic SWNTs (with weak or no gate dependence) exhibit low-bias ($V_{DS}$) conductance in the range of 2 to $4e^2/h$ (for L<1 µm, also see Ref. 7) at room temperature. Upon cooling, the nanotubes with L> 200 nm always show increased conductance. In contrast, the conductance of nanotubes with L<100 nm are largely temperature independent. Fig. 2 shows the conductance vs. gate voltage ($V_{GS}$) curves for a L ~ 60 nm SWNT recorded at several temperatures. The conductance is essentially independent of temperature except for the appearance of conductance oscillations below 40 K (Fig. 2). At 1.5 K, G vs. $V_{GS}$ and $V_{DS}$ evolves into a clear interference pattern (Fig. 2b) that corresponds to a Fabry-Perot resonator.[2] The temperature independence of conductance suggests that electron backscattering by acoustic phonons (twistons)[9] is ineffective in the 60 nm long SWNT and $l_{ap}$(300K) »60 nm. In the low bias regime, transport in SWNTs with L ~ 60 nm is ballistic and free of acoustic phonon scattering even at room temperature.

Transport properties of ohmically contacted metallic SWNTs in the low and high bias regimes have been measured at 290 K as a function of tube lengths. Under low biases, we consistently measure higher conductance for shorter tubes. Current vs. bias voltage ($I_{DS}$ vs. $V_{DS}$) curves in Fig. 3 show that the low bias conductance of SWNTs for a L ~ 55 nm tube is ~3.5 $e^2/h$, higher than ~2 $e^2/h$ for a L ~ 700 nm tube. At high biases, currents saturate at the ~ 20 µA level for long tubes (L ~ 700 nm and 300 nm) due to optical or zone boundary phonon scattering,[1] and reaches 60 µA for a short L ~ 55 nm (Fig. 3). Shorter tubes (L<100 nm) exhibit different slopes in the high-bias regime of the $I_{DS}$-$V_{DS}$ curves than in the low-bias regime, and do not show current saturation.

To determine whether these results were consistent with the expected mean-free-paths, we performed Monte Carlo simulations of the $I_{DS}$-$V_{DS}$ characteristics for metallic SWNTs of various lengths. Monte Carlo simulation solves the Boltzmann transport equation stochastically and has been used extensively for treating high-field transport in semiconductor devices.[10] Carrier transport in one-dimensional metallic tubes differs from that typically encountered in semiconductors since, 1) the metallic band has a linear E-k relation without a band gap, and 2) carriers are highly degenerate, which necessitates the treatment of Pauli blocking of scattering events. The linear E-k relation results in a constant density-of-states and energy-independent scattering rate (in these initial studies, we neglected scattering to the higher, semiconductor like bands). Two scattering mechanisms were included. The first is backscattering by acoustic phonons with an assumed constant scattering rate of $1/\tau_{ap} = v_F/l_{ap}$ where $v_F$ is the Fermi velocity. After backscattering by an acoustic phonon, the carrier energy is conserved and velocity direction reversed. The second is backscattering by optical phonon (with small **k** vector) and zone-boundary phonon (with large **k** vector) emission with a scattering rate of $1/\tau_{op}= v_F/l_{op}$. Only phonon emission was treated; optical and zone-boundary phonon absorption was omitted since the phonon energies are much larger than the thermal energy at room temperature. Elastic scattering due to defects was ignored in the calculations since its mfp is $l_e \geq 1$ μm,[7] greater than $l_{ap}$ for our CVD grown materials. To treat carrier degeneracy, we used an ensemble Monte Carlo method as described by Lugli.[11] The position and energy-dependent distribution function was updated after each time-step so that the probability that a final state was empty could be evaluated. The contacts were assumed to be ideal (i.e. no reflection at contacts). The level of treatment was essentially that of Ref. 1 using the phonon mfps as fitting parameters without treating the detailed phonon dispersion relation (macroscopic theory), but the Monte Carlo simulation facilitates the treatment of so-called off-equilibrium transport in short tubes under high bias. It can also be readily extended to include phonon dispersion characteristics and additional scattering mechanisms.

Fitting the calculated $I_{DS}$-$V_{DS}$ curves to the experimental results reveals a mfp of $l_{ap} \approx 300$nm for acoustic phonon backscattering and a mfp of $l_{op} \approx 15$nm for optical phonon backscattering (Fig. 3). The channel conductance under low bias is controlled by $l_{ap}$ and that under high bias is controlled by $l_{op}$. Notice that the calculation underestimates the measured low bias conductance for L ~ 700nm, so $l_{ap} \approx 300nm$ should be the lower limit of the mfp for acoustic phonon backscattering. The upper limit (which is obtained by fitting the low bias $I_{DS}$-$V_{DS}$ for L ~ 700nm only) may be a factor of ~ 2 larger. The best fitting results to all of the curves are obtained when a phonon energy of $\hbar\Omega \sim 0.2$ eV is used, close to the energies of optical and zone boundary phonon in the range of 1300 to 1580 cm$^{-1}$. Our fitting is nevertheless imperfect, which could be due to errors in the lengths of the short tubes measured by atomic force microscopy, and small variations in the quality of the tubes and contacts for the devices studied. The mfps and optical phonon energies obtained here are similar to those obtained by fitting experimental data [1,12] (Note that Ref 12 parallels with the current work). However, the mfp of $l_{op} \approx 15$ nm is shorter than the ~ 100 nm value that has been estimated from the deformation potential.[1] Further work on the detailed phonon dispersion relations and the deformation potential is needed to understand the difference.

A simple method based on an empirical formulism of resistance can also be used to analyze the experiment $I_{DS}$-$V_{DS}$ curves and extract the phonon scattering mfps. Since optical phonon scattering is dominant over acoustic phonon and elastic scatterings in the high bias regime for short nanotubes, the effective mfp for electron backscattering is largely set by the $l_{op}$, $l_{eff}^{-1} = l_e^{-1} + l_{ap}^{-1} + l_{op}^{-1} \approx l_{op}^{-1}$. The slope of the nanotube $I_{DS}$ vs. $V_{DS}$ probed under high bias can be written as, $G = \Delta I_{DS} / \Delta V_{DS} = G_0 T$, where $G_0 = 4e^2/h$ and $T = l_{op}/(l_{op} + L)$ is the transmission probability[13] between the source and drain in the presence of optical phonon scattering. Fittings of the slopes for all of the $I_{DS}$-$V_{DS}$ curves in the high-bias regime suggest $l_{op} \sim 11$ nm, similar to $l_{op} \sim 15$ nm from Monte Carlo calculations. The results here represent the first time that the strengths of electron couplings with both acoustic and optical phonons are elucidated in a quantitative manner based on length dependent nanotube transport properties.

7The short mfp $l_{op}$ for optical phonon scattering suggests that truly ballistic transport in SWNTs under high bias requires length scaling to the nanometer scale. The shortest metallic SWNT device that we have fabricated is ~ 10 ± 5 nm long between Pd electrodes (uncertainty due to the finite resolution of atomic force microscopy). Up to 70 µA current can flow through the ~ 10 nm tube at $V_{DS}$ ~ 2.6V, beyond which breakdown of the nanotube is observed (Fig. 4a). On the same nanotube, a L ~ 300 nm long segment exhibits a current carrying limit of $I_{DS}$ ~ 25 µA and breakdown voltage of $V_{DS}$ ~ 4 V (Fig. 4b), while a L ~ 3 µm long segment breaks down at a much higher voltage of $V_{DS}$ ~ 13.8 V and lower current of $I_{DS}$ ~ 18 µA (Fig. 4c). In shorter nanotubes, electrons encounter less backscattering and accelerate to high energies rapidly, and thus breakdowns occur at lower biases. For the 3 µm long nanotube, electrical breakdown most likely occurs at defect sites on SWNTs (the mfp for elastic defect elastic scattering is typically $l_e$< 3µm for CVD grown nanotubes[5,7]).[14] For the medium length L ~ 300 nm nanotube section, breakdown may be related to joule heating as a result of electron/optical phonon coupling and the resulting oxidation in air.[15] The breakdown of the ultra short L ~ 10 nm SWNT segment appears different from the longer segments in that its $I_{DS}$-$V_{DS}$ curve exhibits only slight flattening beyond $V_{DS}$ ~ 0.2 V (indicating optical phonon scattering not as effective), and shows a steeper slope beyond ~ 1.4 V prior to breakdown at ~ 2.6 V. The steeper $I_{DS}$-$V_{DS}$ beyond ~1.4 V is likely to be due to conduction through the first non-crossing higher sub-band in the metallic tube, which has a band gap of $E_g \approx 2.6(eV)/d(in\,nm) \approx 1.3 eV$ .[16] The mechanism of eventual breakdown could involve high field (~ $10^6$ V/cm) impact ionization[17,18] (likely assisted by residue optical phonon scattering) involving carriers in the higher subbands. It is also possible that

electro-migration occurs in the thin and small Pd electrodes and metal-tube contacts degrade under the high current density (~$10^7 A/cm^2$ in the Pd electrodes). Due to the importance of high current delivery capability of SWNT devices, we have also investigated high field transport in Pd ohmically contacted short (L ~ 50 nm, slightly longer than $l_{op}$) semiconducting SWNTs (s-SWNT).[5] In the p-type on-states, short s-SWNTs exhibit an upturn beyond $V_{DS}$ ~ 1 V prior to complete breakdown at ~ 2.2 V (maximum current ~ 70 μA) (Fig. 4d). In this case, the current upturn can be attributed to injections of electrons to the conduction band[19] and subsequently into higher subbands of the nanotube under the increasing bias voltage. The breakdown of semiconducting nanotubes could be due similar effects as in metallic nanotubes as described above. Detailed work is necessary to fully understand transport near the breakdown regime in quasi-ballistic nanotubes.

In summary, SWNTs with highly reproducible ohmic contacts and various lengths are investigated by electron transport experiments and theory. The mean free paths for acoustic and optical phonon scatterings are determined. Short nanotubes (~ 10 nm) are ballistic and quasi-ballistic in the low and high bias regimes respectively. Remarkably high currents up to 70 μA can flow through these short tubes. The eventual breakdowns for ballistic metallic and semiconducting nanotubes are discussed and deserve further investigation due to their importance to high power transistors or interconnects.

The authors are indebted to Prof. Supriyo Datta of Purdue University and Dr. M. P. Anantram of NASA Ames Research Center for extensive technical discussions. This work was supported by MARCO MSD Focus Center, DARPA's Moletronics program, and the NSF Network for Computational Nanotechnology.

**Figure Captions**

**Fig 1.** Atomic force microscropy (AFM) images of five devices consisting of individual SWNTs with lengths in the range of L=600 nm to 10 nm between the edges of Pd contact electrodes.

**Fig. 2.** Electrical properties of a L ~ 60 nm long ohmically contacted metallic SWNT (diameter ~ 1.5 nm). a) G vs.$V_{GS}$ recorded (under a low bias of $V_{DS}$=1 mV) at 290 K, 150 K and 40 K respectively. A noteworthy technical point is that thin Pd contact electrode fingers were used (250 nm × 25 nm) for contacts in order to form small (< 100 nm) gaps between electrodes by liftoff. The resulting narrow Pd electrodes had a resistance of ~ 0.6 kΩ at 290K and a negligibly small resistance of 0.019 kΩ at 40 K. The series resistance was estimated from the geometry of the contact electrodes and the resistivity of the Pd film at various temperatures. The data shown here was after the correction of the series resistance of the Pd electrodes. b) Differential conductance vs. $V_{DS}$ and $V_{GS}$ for the device recorded at 2 K. The arrow points to a resonance peak,[2,7] at $V_{DS} = \pi\hbar v_F/eL$ = 25 mV where $v_F = 8.1\times10^5$ m/s is the Fermi's velocity and L ~ 60 nm. The gate oxide thickness $t_{ox}$ ~ 10 nm for the device in this figure.

**Fig. 3.** Electrical properties of ohmically contacted metallic SWNTs of various lengths (diameters d~2-2.5 nm, oxide thickness $t_{ox}$ ~ 10 nm for the samples in this figure). Solid lines are experimental $I_{DS}$-$V_{DS}$ curves and the symbols are Monte Carlo calculation and fitting results. Note that the devices in Fig. 3 had a series resistance of ~ 1 kΩ arising from the thin Pd electrodes except for the 55 nm long tube which had an electrode

resistance of ~ 3 k$\Omega$. The experimental curves are all after correction of $V_{DS}$. We also note that for short tubes ( < 100 nm), a permanent change in the electrical properties was observed after applying large fields ($V_{DS}$> 1V). This could be due to the degradation of the contacts at such high drain biases, and requires further work to illuminate the exact cause.

**Fig. 4.** $I_{DS}$-$V_{DS}$ breakdown characteristics of SWNTs. The lengths of three sections of a metallic SWNT are a) ~ 10 nm, b) ~ 300 nm and c) ~ 3 µm, with the Si substrate being grounded. The gate oxide thickness $t_{ox}$ ~ 67 nm for the samples. d) $I_{DS}$-$V_{DS}$ data recorded for a semiconducting SWNT (in the ON state under $V_{GS}$=0V) with L ~ 50 nm and $t_{ox}$ ~ 10 nm (inset shows $I_{DS}$ vs. $V_{GS}$ under $V_{DS}$=10 mV). AFM image insets in a)-c) show the devices after electrical breakdown, and arrows in b) and c) point to the breakdown points. The parasitic electrode resistance is high (~ 15 k$\Omega$, due to the very thin and narrow electrode geometry) for devices in a)-c). For this reason, they were not used for the quantitative analysis of $l_{op}$ in Fig. 3.

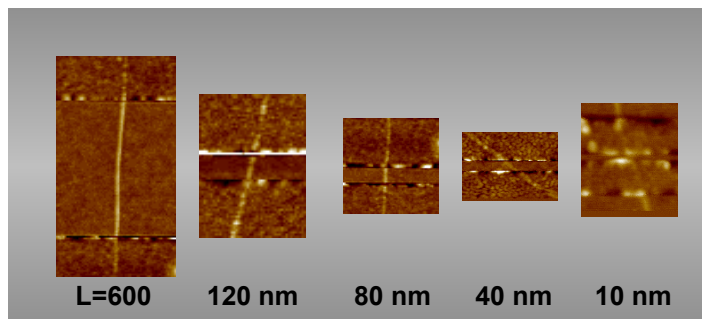

**Figure 1**

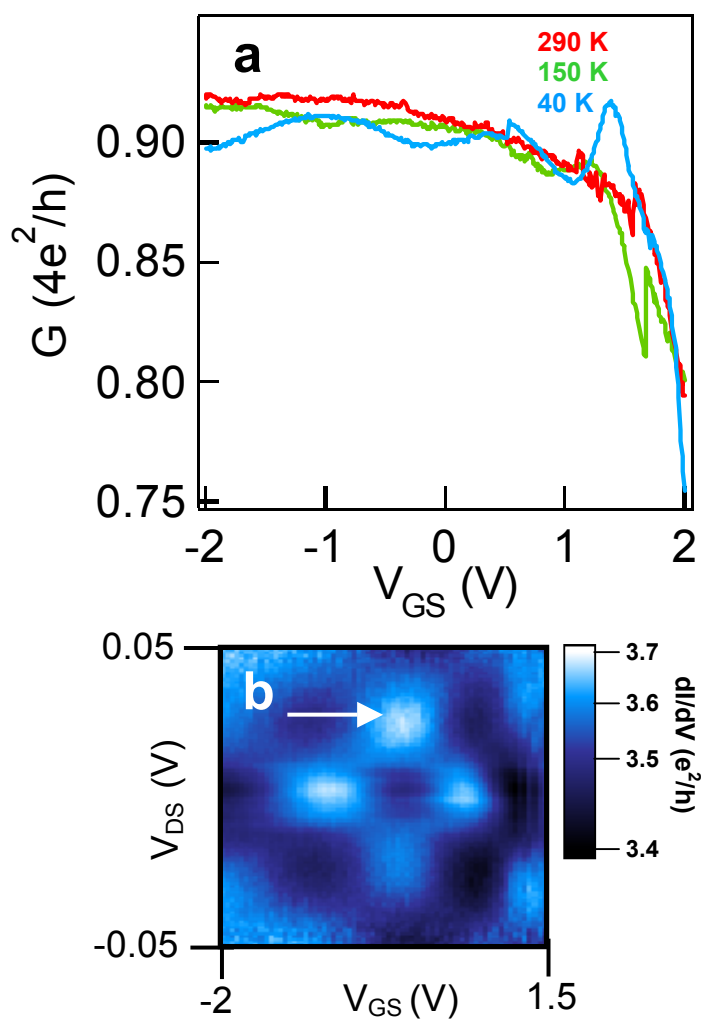

**Figure 2**

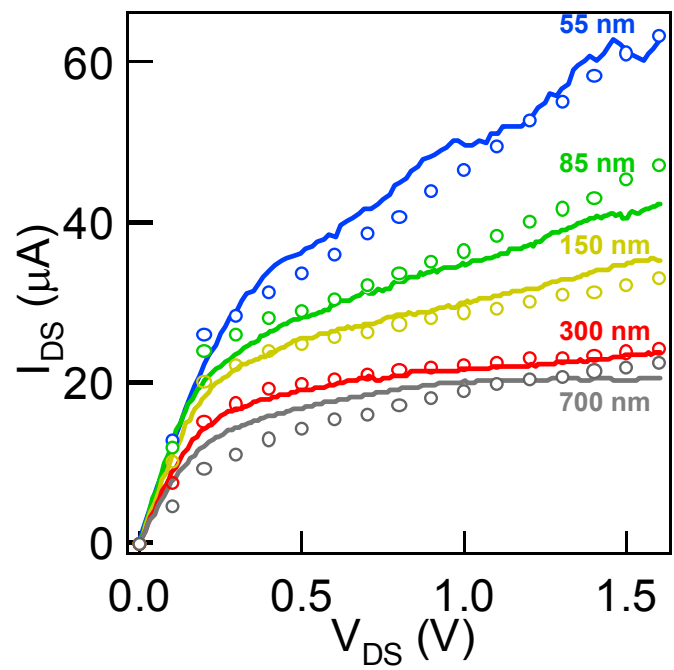

**Figure 3**

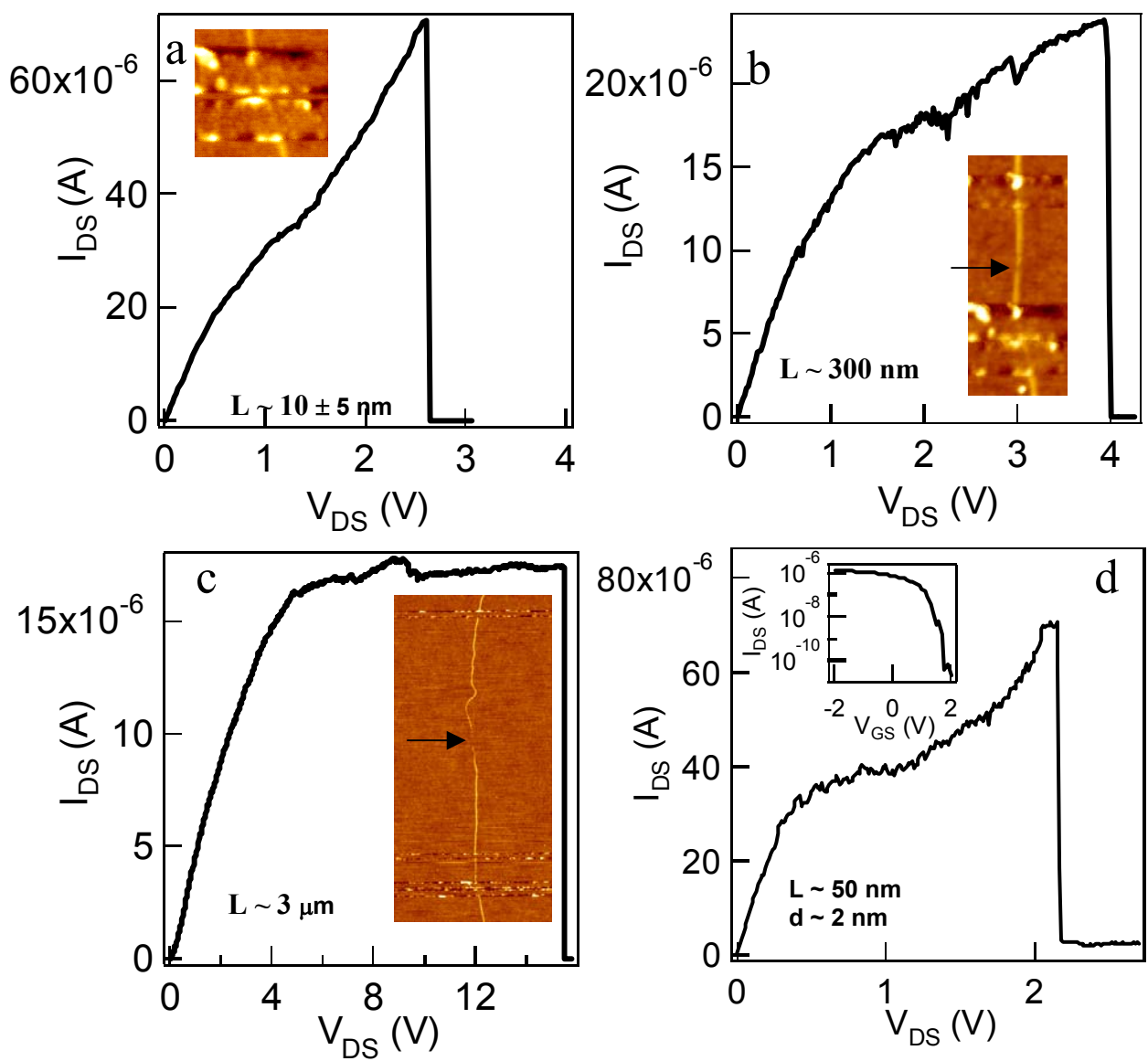

**Figure 4**